\documentclass[12pt,journal,draftclsnofoot,onecolumn]{IEEEtran}
\usepackage{amsfonts}
\usepackage{amssymb}
\usepackage{amsmath}
\usepackage[dvips]{graphicx}
\usepackage{cite}

\usepackage{enumitem}
\usepackage{balance}
\usepackage{multirow}
\usepackage{balance}
\usepackage{fixmath}
\usepackage{balance}
 \usepackage{algpseudocode}
  \usepackage{algorithm}
\begin{document}
\title{Low Power Wide Area Networks (LPWANs) for Internet of Things (IoT) Applications: Research Challenges and Future Trends}

\author{ Alexandros~--~Apostolos~A.~Boulogeorgos,~\IEEEmembership{Member,~IEEE}, \\ Panagiotis D. Diamantoulakis,~\IEEEmembership{Student Member,~IEEE},  
and \\ George K. Karagiannidis, \IEEEmembership{Fellow, IEEE} 

\thanks{The authors are with the Department of Electrical and Computer Engineering, Aristotle University of Thessaloniki, 54 124, Thessaloniki, Greece (e-mail:{ 
\{ampoulog, pdiamant, geokarag\}@auth.gr)}.
}
}
\maketitle	

\begin{abstract}

Internet of things (IoT) changes significantly the requirements for connectivity, mainly with regards to long battery life, low device cost, low deployment cost, extended coverage and support for a massive number of devices. 
Driven from these requirements, several different cellular and non-cellular low power wide area network (LPWAN) solutions are emerging and  competing for IoT business and the overall connectivity market. Motivated by this, in this paper, we review and compare the design specifications of different LPWAN solutions, as well as, we discuss their suitability for different IoT applications. Finally, we present the challenges, future trends, and potential  research directions for LPWANs.

\end{abstract}


\section{Introduction}\label{S:Intro}

\begin{figure}
\centering\includegraphics[width=0.95\linewidth,trim=0 0 0 0,clip=false]{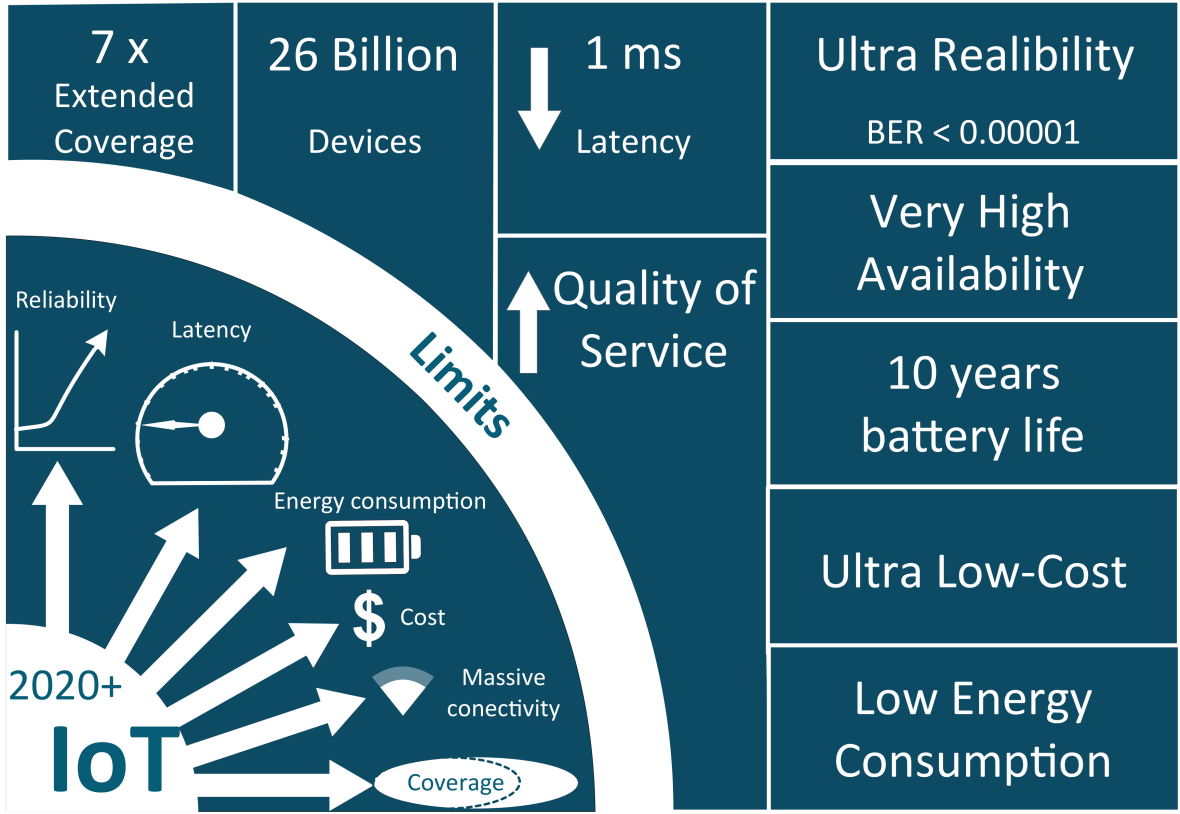}
\caption{IoT requirements and challenges.}
\label{fig:IoT}
\end{figure}

During the last decades, wireless communications have been a subject of much hype, due to their increasing integration in everyday life. As a result, they have evolved significantly from early voice systems to today's highly sophisticated integrated communication platforms that provide numerous services, which are used by billions of people around the world. The internet of things (IoT) is considered as the next revolution of communications, which will play a significant role in improving the  efficiency of the human, natural and/or energy resource management, as well as in optimizing the production processes. 
As a consequence, by 2020, it is expected that approximately $26$ billion IoT devices will serve the global population~\cite{IoT_26_Billions}.
On the other hand, as illustrated in Fig.~\ref{fig:IoT}, as IoT systems evolve, we are comforting more and more inherent  limitations, preventing further performance improvements.  
Therefore, it is necessary to develop the appropriate technologies, which meet those requirements.  

Since recently, there was no economical, flexible, and reliable technology for the connection of the IoT devices in the network. 
Traditional solutions, such as short-range wireless networks, e.g. Bluetooth, ZigBee, Z-Wave, wireless local area networks (WLANs), e.g. wireless fidelity (WiFi), HiperLAN, and cellular networks, e.g., global system for mobile communications (GSM), long-term evolution (LTE), etc., even though they allow the wireless connection of the IoT devices in the network, they are usually of high cost, high energy consumption, high complexity and low reliability approaches.   
As a result, the technology of low power wide area networks (LPWANs) has been recently developed. 
LPWANs are considered excellent candidates for IoT applications, since they promise high energy efficiency, low power consumption and high coverage capabilities. 

The major industrial representatives of LPWANs are Semtech, Sigfox, and Huwai, with LoRA, ultra narrow-band (UNB), and narrow-band (NB)-IoT approaches, respectively. 
Additionally, three separate tracks for licensed cellular IoT technologies are  standardized by the third generation partnership project (3GPP), namely 
enhanced machine type communication (eMTC), often referred to as LTE-M~\cite{WP:LTE_evolution_for_IoT_connectivity}, and extended coverage GSM (EC-GSM).
Finally, a fifth generation (5G) solution for cellular IoT is expected to be part of the new 5G framework by 2020~\cite{WP:LTE_M_Optimizing_LTE_for_the_IoT}. 

Motivated by the IoT applications requirements and the new technologies that emerged in order to deal with them, in this article, we first review and compare the design specifications of the different LPWAN approaches.  Furthermore, we revisit the IoT applications and, based on the specifications of the LPWANs, we discuss the suitability of each LPWAN  to meet the requirements of each IoT application. Finally, we present the challenges and future trends of LPWANs, as well as possible technological solutions and the potential research~directions.   

The rest of this paper is organized as follows. In Section~\ref{S:Specifications}, the specifications of the different LPWAN technologies are reviewed, whereas in Section~\ref{S:Suitability}, based on the requirements of the IoT applications and the specifications of LPWANs, we discuss the suitability of the different LPWAN technologies for the IoT applications. 
Likewise, in Section~\ref{S:Challenges}, the challenges and the future trends for LPWANs are presented, while closing remarks are provided in Section~\ref{S:Conclusions}.

\section{Requirements and design specifications of LPWANs}\label{S:Specifications}

Numerous applications are envisioned for IoT, including utility meters, vending machines, automotive, metering and alerting. 
The key requirements for LPWANs, which can successfully support those IoT applications are:
\begin{itemize}
\item \textit{High energy autonomy of the connected devices}. Several IoT applications require long battery life devices, i.e., the IoT devices should be able to operate without battery replacement for a long period of time. An indicate example is the use of fire sensors, which alarm fire departments in  case of fire. In such services, the battery life is a significant cost factor, whereas changing batteries in short time periods might not be feasible. 
\item \textit{Low device and deployment costs.} In order to enable a profitable business case for IoT, the total cost of device's ownership should be extremely low. As a result, the current industry targets in a modulo cost of less than $5\$$~\cite{WP:Cellular_networks_for_massive_IoT}. Moreover, the deployment cost, including the initial capital expenditures and annual operating expenses should be kept to minimum, by simple upgrading the existing infrastructures and avoiding the deployment of new hardware~\cite{WP:LTE_evolution_for_IoT_connectivity}. 
\item \textit{Extended coverage.} LPWANs target in enabling deeper indoor coverage by enhancing the IoT connectivity link budget for $15-20\text{ }\mathrm{dB}$.
This requirement is necessary in order to support IoT devices, such as smart meters, which are located deep indoor, e.g. in the basement of buildings, behind concrete walls, inside elevators, etc.  
\item \textit{Support for a massive number of devices.} Due to the exponential increase of the IoT devices, which may not be uniformally distributed, each LPWAN base-station (BS) should be able to support a massive number of simultaneously connected IoT devices. 
\end{itemize}

Based on those requirements, the basic design specifications of LoRA, UNB, NB-IoT, LTE-M, and EC-GSM LPWANs are summarized in Table~\ref{T:LPWANs}~\cite{WP:LTE_evolution_for_IoT_connectivity,WP:LTE_M_Optimizing_LTE_for_the_IoT,TR:The_way_to_the_connected_planet,TR:Cellular_system_support_for_ultra_low_complecity_and_low_throughput_IoT,TR:NB_IoT,
WP:LoRA}. From this table, it is evident that LPWANs achieve long range and low power operation at the expense of low data rate and higher latency (typically in orders of seconds or~minutes).

\begin{table*}
\centering%
\renewcommand{\arraystretch}{1.3}
\caption{The basic specifications of LPWANs.}
\label{T:LPWANs}
\begin{tabular}{|c||c|c|c|c|c|c|c|c|}
\hline
\textbf{} & \normalsize{\textbf{LoRA}} 	&  \normalsize{ \textbf{UNB}} &  \normalsize{\textbf{NB-IoT}} &  \normalsize{\textbf{LTE-M}} &\normalsize{\textbf{LTE-M}}    & \normalsize{\textbf{EC-GSM}} \\
\textbf{} &  					  	&                        &                           & \normalsize{\textbf{Rel. 12/13}} &\normalsize{\textbf{Rel. 13}}& \normalsize{\textbf{Rel. 13}}   \\
\hline \hline
{\textbf{Range}} &  $< 11\text{ }\mathrm{km}$ 	&  $< 13\text{ }\mathrm{km}$ & $< 35\text{ }\mathrm{km}$	&	$<11\text{ }\mathrm{km}$ &	$<15\text{ }\mathrm{km}$ &			$<15\text{ }\mathrm{km}$\\
\hline
{\textbf{Max. coupling loss}} & $157\text{ }\mathrm{dB}$	 & $160\text{ }\mathrm{dB}$&$164\text{ }\mathrm{dB}$& $156\text{ }\mathrm{dB}$ & $164\text{ }\mathrm{dB}$& $164\text{ }\mathrm{dB}$\\
\hline
{\textbf{Spectrum}}  & Unlicensed 	&Unlicensed	& Licensed&	 Licensed	&	Licensed &Licensed \\
{\textbf{Bandwidth}} & $900\text{ }\mathrm{MHz}$ & $900\text{ }\mathrm{MHz}$ &	$7-900\text{ }\mathrm{MHz}$ & 	$7-900\text{ }\mathrm{MHz}$	&$7-900\text{ }\mathrm{MHz}$	& $8-900\text{ }\mathrm{MHz}$  \\
		&  $<500\text{ }\mathrm{kHz}$		   &	$100\text{ }\mathrm{Hz}$ & $200\text{ }\mathrm{kHz}$ & $1.4\text{ }\mathrm{MHz}$	or & $200\text{ }\mathrm{kHz}$	or& $2.4\text{ }\mathrm{MHz}$	or		   \\
						   &  		   &	 &  &		shared		& shared	& 	shared\\
\hline
e{\textbf{Data rate}} & $10\text{ }\mathrm{Kbps}$ &$<100\text{ }\mathrm{bps}$&	$<170\text{ }\mathrm{Kbps}\text{ }$(DL)&	$<1\text{ }\mathrm{Mbps}$	&	$<150\text{ }\mathrm{Kbps}$	 &	$<10\text{ }\mathrm{Kbps}$ \\
						 &  & &	$<250\text{ }\mathrm{Kbps}\text{ }$(UL)&						  &										   & \\
\hline
{\textbf{Indoor}} & Yes & No & Yes & No & Yes & Yes\\ \hline
{\textbf{Security}} & No & No & Yes & Yes & Yes & Yes \\ \hline
{\textbf{Bi-directional}} & Yes & Yes & Yes & Yes & Yes & Yes  \\  \hline
{\textbf{Global ecosystem}} & No & No & Yes & Yes & Yes & Yes  \\ \hline
{\textbf{Battery life}}& $>10\text{ }$years &	$>10\text{ }$years &$>10\text{ }$years	&	$>10\text{ }$years	&	$>10\text{ }$years	 &		$>10\text{ }$years							   \\
\hline
\end{tabular}
\end{table*}

LoRA and UNB are both proprietary solutions, which are deployed in the unlicensed industrial, scientific and medical (ISM) radio bands. Therefore, they are not globally supported. 
On the other hand, NB-IoT, LTE-M, and EC-GSM are licensed cellular IoT solutions that are standardized be the third generation partnership project (3GPP). In particular, NB-IoT offers three deployment scenarios, namely stand alone, guard-band, and in-band~\cite{WP:Huawei_NB_IoT}.  
In the stand alone deployment, underutilized bandwidth is used, while in the guard-band deployment allocated bandwidth, which is not utilized by LTE carriers, is used. In-band NB-IoT is deployed in LTE assigned carriers.

The UNB, even though it provides the lowest data rates, it does not support indoor communications and  it does not guarantee security and privacy of the transmitted data. Nevertheless, since it was one of the first deployed commercial LPWAN, it already holds a great market share in the western Europe. 
On the other hand, EC-GSM provides data rates up to $10\text{ }\mathrm{Kbps}$, it supports indoor communications and it employs encryption techniques to ensure security and privacy.  
Similarly, LoRA achieves data rates up to $10\text{ }\mathrm{Kbps}$ and it can be employed for indoor communications; however, in the contrary to EC-GSM, it has no security mechanism.     
LTE-M release 13 and NB-IoT provides medium data rates and security schemes, as well as it can be used for indoor communications. Finally, LTE-M release 12/13 provides the maximum data rates, it uses security mechanisms; however, it does not support indoor~communications.  

\section{LPWANs for IoT Applications}\label{S:Suitability}

There will be a wide range of IoT applications in the future, and the market is now expanding toward  massive IoT deployment, as well as more advanced solutions that may be categorized as critical IoT~\cite{WP:Cellular_networks_for_massive_IoT}. Critical IoT applications refer to remote health care, traffic safety and control, smart grid control applications, complex industrial applications, as well as manufacturing, training, and surgery.
Therefore, they have very high demands for reliability, availability and low latency.
On the other hand, massive IoT  typically  consists of sensors that report to the cloud in a regular basis. Hence, massive IoT demands  low-cost devices with low energy consumption and good coverage. In other words, LPWANs are suitable for massive IoT applications. 

\begin{figure}
\centering\includegraphics[width=0.95\linewidth,trim=0 0 0 0,clip=false]{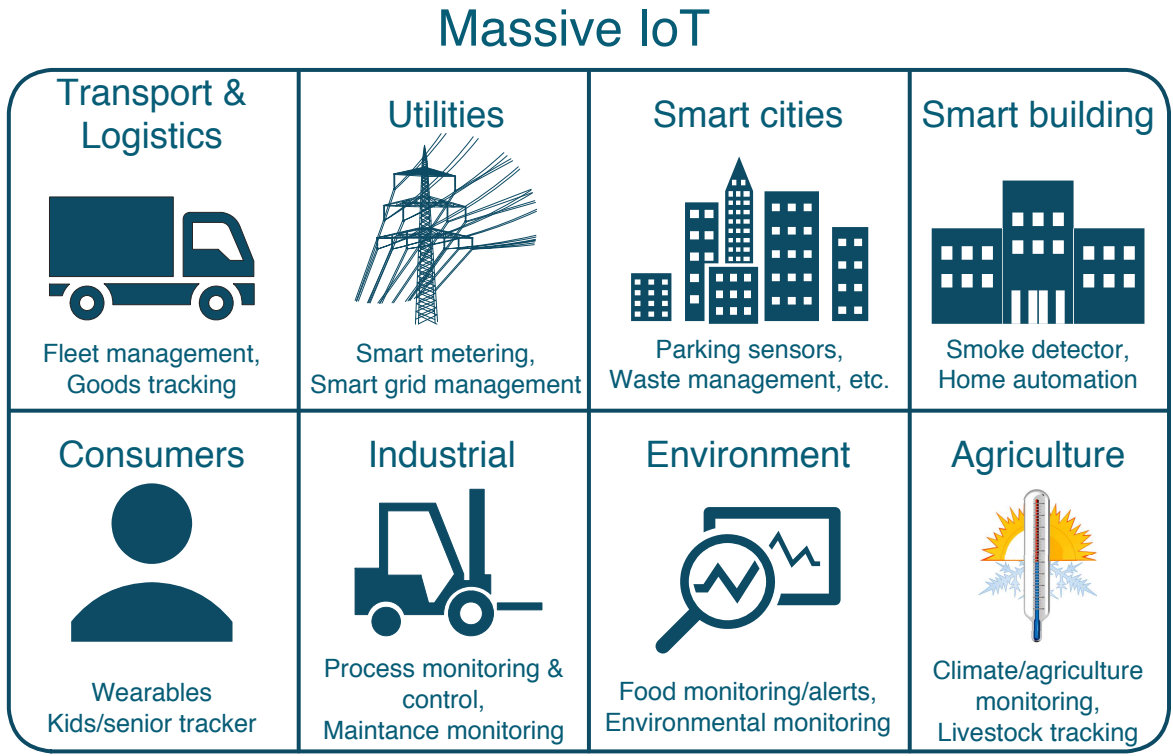}
\caption{Massive IoT applications enabled by LPWANs.}
\label{fig:Massive_IoT}
\end{figure}

\begin{figure}
\centering\includegraphics[width=0.95\linewidth,trim=0 0 0 0,clip=false]{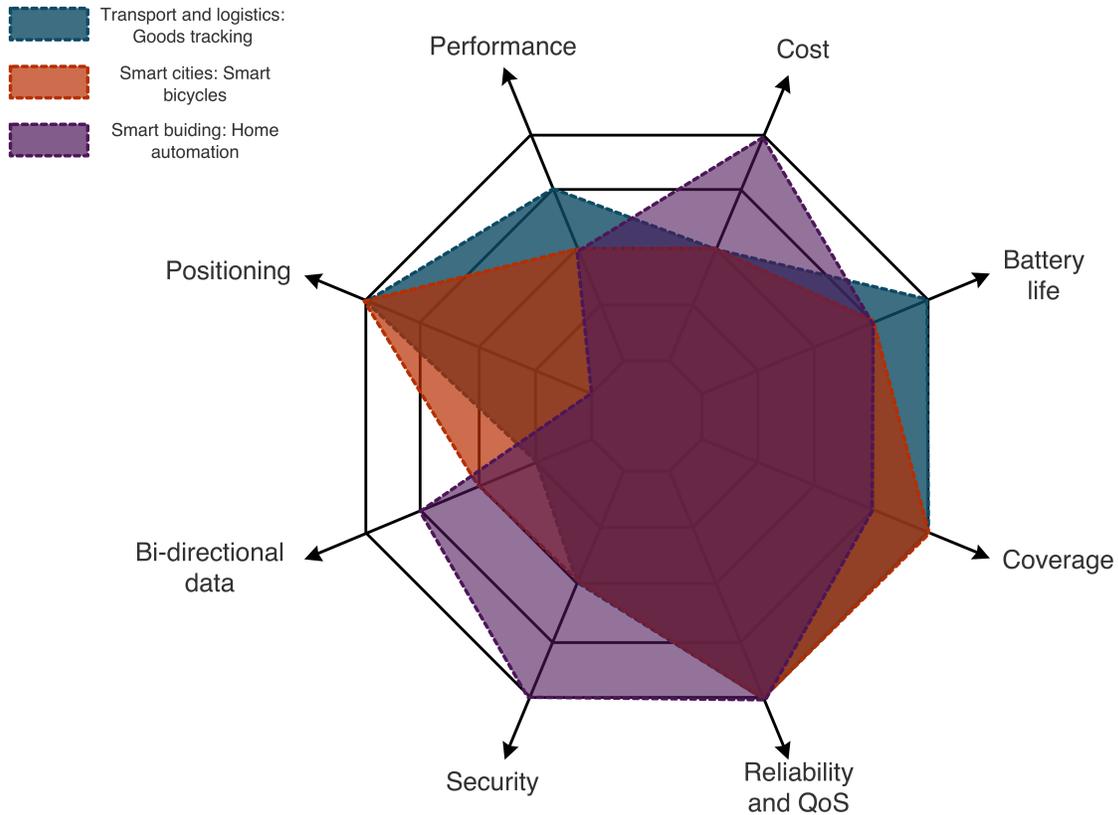}
\caption {Radar chart showing indicative IoT applications requirements~\cite{WP:Cellular_networks_for_massive_IoT}.}
\label{fig:Applications_IoT_Requirements}
\end{figure}

As illustrated in Fig.~\ref{fig:Massive_IoT},  massive IoT market includes several widely used applications in several sector, such as transports \& logistics, utilities, smart cities, smart buildings, consumers electronics, industry, environment, and agriculture. As a result, massive IoT applications have a huge  variety of requirements
regarding cost, battery life, coverage, connectivity performance (throughput and capacity), security and reliability.  For instance, as demonstrated in Fig~\ref{fig:Applications_IoT_Requirements}, security is important for home automation IoT applications, while it is not a key requirements for goods tracking applications. This is because in applications, such as building security, sensitive information could be broadcasted, which demands strict security. Furthermore, in the case of a break-in, it is crucial that the alarm information reaches the control center within time-making duration; as a consequence, two-way communication is vital for those~applications. 

Another crucial parameter for IoT applications is throughput. Some applications requires from the end devices to send a few messages per day, while other requires to transmit a large amount of data. For example, status indicators for temperature will send a very small amount of data, while camera sensor that transmit a video stream to guide a remote repair technician or used for security check will send a much greater amount of data. 
Consequently, the difference in throughput requirements is huge. 
Thus, operators or service providers, which handle several applications, should harmonize communication modules, so that they all use the same
underlying radio solution to reduce operational and fault management cost and complexity, while, at the same time, they should carefully select the appropriate technology, which will meet the requirements of the services that they offer. 

Several applications require both monitoring and control of remote devices. For instance, an IoT application for green houses, which monitors the temperature  and decides whether to increase or decrease it by using a heating or ventilation/cooling mechanism, respectively. 
From the technology point of view, these applications require two-way communications, i.e., both downlink and uplink should be enabled. 
Moreover, two-way communications enables simple software updates of the remote IoT devices, and increase the reliability of the link, as well as the fault management of the network, since the LPWAN nodes (both the BSs and the IoT devices) can send acknowledgment for the received~messages.   

\begin{figure}
\centering\includegraphics[width=0.95\linewidth,trim=0 0 0 0,clip=false]{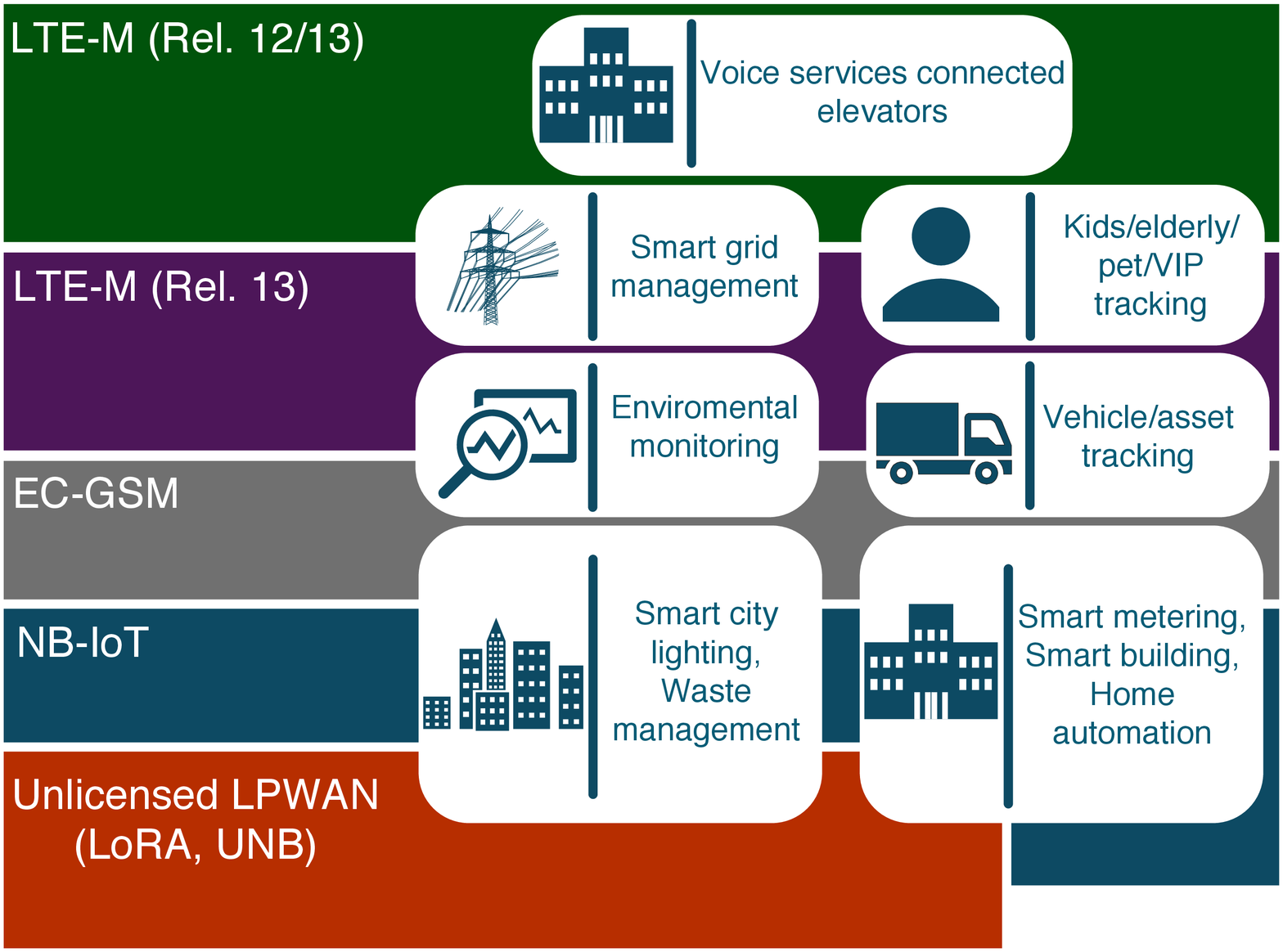}
\caption{Suitability of different LPWANs for different IoT applications.}
\label{fig:Applications_IoT}
\end{figure}

All the above mentioned IoT applications indicate that different LPWANs are expected to be used to meet the different requirements of the different services. Fig.~\ref{fig:Applications_IoT} matches applications with the appropriate LPWANs that can meet their requirements.
According to this figure, unlicensed LPWANs, such as LoRA and UNB, can support local, low data rate demanding smart cities and building applications. 
Specifically, LoRA and UNB can support lighting control and waste management operations, whereas LoRA can additionally support low data rate smart building and home automation applications. On the other hand, UNB is capable of indoor communications; therefore, it cannot meet the requirements of smart building operations. 
Likewise, more data rate hungry smart cities and smart building applications should employ NB-IoT. 

IoT applications that need to be deployed in a global scale can use EC-GSM, LTE-M release 13, or LTE-M release 12/13. The selection of the appropriate licensed LPWAN depends on the data rates and reliability demands. For instance, environmental monitor and vehicle or assets tracking applications can use either EC-GSM or LTE-M release 13, depending on the network availability, the data rate demands and the required reliability. On the other hand, smart grid management and kids/elderly/pet/very important person (VIP) tracking application can be supported by LTE-M release 13, which meets the necessary data rate and security  demands.     
Finally, LTE-M release 12/13 can be employed in order to implement applications, in which reliability and high data rates are key requirements. 

\section{Research Challenges and future trends}\label{S:Challenges}

\begin{figure}
\centering\includegraphics[width=0.85\linewidth,trim=0 0 0 0,clip=false]{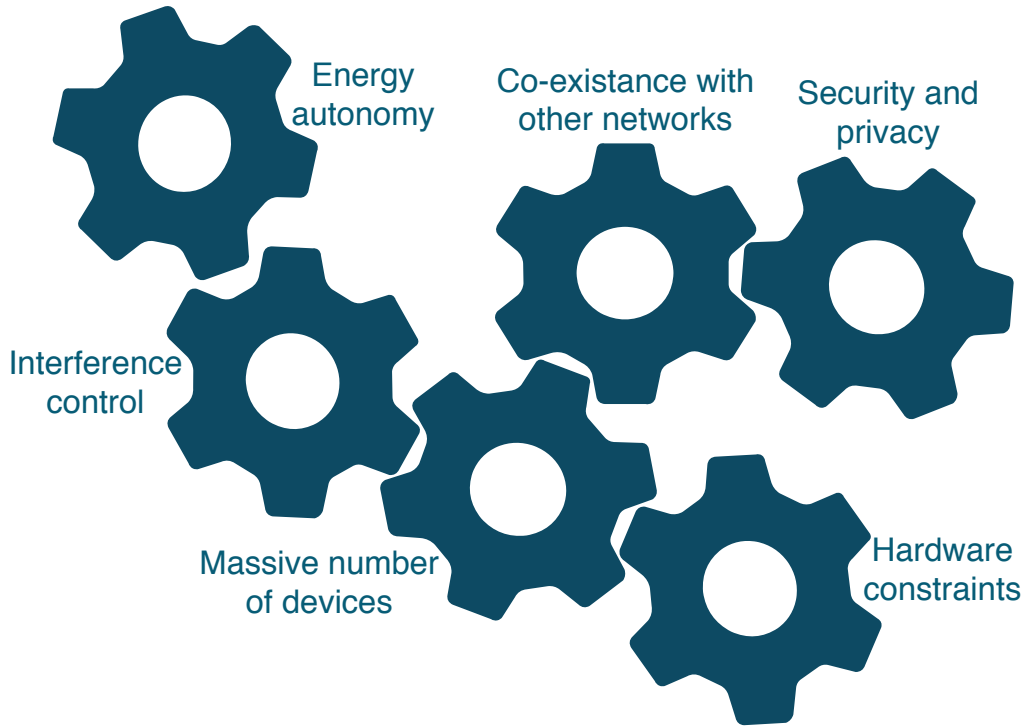}
\caption{Challenges in LPWANs.}
\label{fig:Challenges}
\end{figure}

LPWAN companies strive hard to innovate solution that can deliver low-power consumption, low-cost, reliable and high performance services. In this race, it is easy, but counter-productive, to overlook important challenges faced for LPWANs, such as the necessity to support a massive number of IoT devices, interference issues, due to the co-existence with other wireless networks, hardware constraints and performance degradation, due to the low-cost deployed IoT devices, the challenge of 10-years energy autonomy devices, as well as security issues. These challenges are graphically illustrated in Fig.~\ref{fig:Challenges}.  

\subsection{Support a huge number of devices}

LPWANs are expected to connect ten of millions of IoT devices, which will transmit data over confined and often shared radio resources. The resource allocation problem is further burdened by the device density across different geographical areas, as well as the cross technology interference. Especially, the unlicensed LPWAN technologies may suffer from dramatic performance degradation, due to interference from single broadband signals. Likewise, licensed LPWANs, which operate in-band with voice and video services, are also at this risk. Finally, most LPWAN solutions employ medium access control (MAC) protocols, such as ALOHA and carrier sense multiple access (CSMA), which, although they have low complexity, they do not scale efficiently with the increasing number of the connected IoT devices~\cite{A:Goodbye_ALOHA}.    

Several research directions can be pursued to address the resources issues of LPWANs. These include the use of channel diversity, opportunistic spectrum sensing, adaptive transmission strategies, non-orthogonal multiple access schemes (NOMA).  The use of multi-mode multiple antennas at the LPWANs BSs can significantly increase the diversity gain or the data rates of specific IoT devices, located in the coverage area of the BS. Furthermore, it can enable space division multiplexing, which can boost the number of IoT devices that the BS can support. 

Opportunistic spectrum access can aid in improving the spectrum efficiency, by exploiting the under-utilized bandwidth. Specifically, IoT devices can perform spectrum sensing and use spectrum holes in order to connect to the LPWAN. This will increase the utilization of the spectrum, as well as it will contribute in the avoidance of self- and cross-interference. 

Adaptive transmission strategies that take into account the peculiar traffic pattern of IoT devices, should be produced in order to mitigate the impact of cross-technology interference. Those techniques will enable the dynamic orthogonalization of LPWANs with the already existing networks in a specific coverage area. In other words, they will render the efficient scale of the LPWAN with the increase of the IoT devices.    

Moreover, to cater areas with high IoT device density, the use of NOMA-based schemes promises be constructive. NOMA, which was proved to offer considerable gain in terms of spectral efficiency and outage probability, while better utilizing the heterogeneity of channel conditions, has been recently proposed for LTE-A, where
it is termed as multi-user superposition transmission (MUST). Also, it has been recognized as
a promising multiple access technique for fifth generation (5G) networks \cite{DOCOMO}. This is also because NOMA can be applied to diverse QoS requirements, i.e., it is appropriate to support the connection of a great number of devices and sensors that require different target rates, such as the ones used in IoT applications. NOMA is fundamentally different from orthognal multiple access schemes, since its basic principle is
that the users can achieve multiple access by exploiting the power domain. In non-orthogonal
techniques, all the users can utilize resources simultaneously, which lead to inter-user interference, which calls for multi-user detection (MUD) techniques to retrieve the users' signals at the receiver, such as joint decoding or successive interference cancellation. It must be noted that the implementation of uplink NOMA, in contrast to downlink NOMA, is not a burden for the users, i.e., the
encoding complexity at the nodes’ side is not affected, since MUD is only
applied at the access point. 

Finally, improvements of the existing MAC protocols that takes into accounts the characteristics of LPWANs and the massive devices demands are required. 
 
\subsection{Interference mitigation}

In the heterogeneous environments of tens of wireless technologies and massive number of devices, which all share the same radio resources, transceivers undergo inevitable high levels of cross- and self-technology interference, which causes significant degradation to the quality of the wireless link. In particular, unlicensed LPWANs, such as LoRA and UNB, which are deployed in the ISM bands and use the ALOHA scheme are very sensitive to those types of interference, since ALOHA enables data transmission of a specific device without requiring the knowledge of the transmission state of other devices. 

In order to enhance the performance of LPWANs, interference issues should be addressed. In this context, low-complexity and high energy efficiency interference mitigation techniques should be borrowed from  traditional cellular networks, as well as new novel interference compensation techniques should be proposed, which takes into accounts the IoT devices capabilities. Furthermore, scheduling schemes should be adopted, in order to suppress the effect of self-technology~interference. 
 
\subsection{Hardware complexity}

From a technology point of view, in order to achieve the required adaptability of the IoT devices, radio transceivers need to be flexible and software-reconfigurable devices. By definition, flexible radios are characterized by the ability to operate over multiple-frequency bands, and to support different types of waveforms, as well as various air interface technologies of currently existing and emerging wireless systems. The flexibility of transceivers, in line with the software define radio (SDR) principle, will enable the use of emerging LPWAN solutions and waveforms though software updates, without hardware changes. Moreover, SDR is considered one of the key technologies that enables the use of opportunistic spectrum access~\cite{A:Energy_detection_spectrum_sensing_under_RF_imperfections}.

From an economical point of view, the advantages in integrated circuit (IC) technologies and the adoption of low-complexity transceiver structures, such as the direct-conversion radio (DCR) architecture, allowed improvements in manufacturing efficiency and automation that resulted in reducing the cost-per-device. Moreover, the use of low-complexity transceiver structure enable the reduction of the power consumption in battery-powered devices, without sacrificing too much~performance. 

In general, the demands for flexibility, as well as the constraints of product cost, device size, and energy efficiency, lead to the use of simplified radio architectures and low-cost radio electronics. In this context, the DCR architecture of such systems provides an attractive front-end solution for LPWANs, as it requires neither external intermediate frequency filters nor image rejection filter. However, these transceivers suffer for radio frequency imperfections, such as in-phase and quadrature imbalance, phase noise, and amplifier's nonlinearities. 
In order to mitigate the negative impact of these imperfections, several digital processing techniques have been proposed.  These approaches usually require high complexity processing, which is energy demanding. Therefore, low-complexity RF impairments mitigation techniques need to be investigated for LPWANs.  

\subsection{Energy autonomy of the connected devices}
Reliable and uninterrupted operation of IoT applications is limited by the finite battery capacity of the utilized devices. 
Nowadays, IoT devices can last no more than 2-3 years. However, until 2020, LPWANs are expected to prolong the battery life of the end devices in order to surpass 10 years. One possible solution that can be used is energy harvesting (EH), which refers to harnessing energy from
the environment or other energy sources and converting to electrical energy, is regarded
as a disruptive technological paradigm to prolong the lifetime of energy constrained
wireless networks. Apart from offering a promising solution for energy sustainability
of wireless nodes in communication networks, EH also reduces
considerably the operational expenses~\cite{Sude}. 

However, the main disadvantage of traditional energy harvesting methods is that they rely on natural resources,
such as solar and wind energy, which are uncontrollable. For
this reason, harvesting energy from radio frequency signals,
which also transfer information, seems to be an interesting
alternative. This technique, commonly known as simultaneous wireless information and
power transfer (SWIPT) is the very challenging challenging, as it presupposes the efficient design
of systems. Note that in practice nodes cannot harvest energy and receive/transmit information simultaneously. In order to overcome this difficulty, two strategies have been proposed, i.e. \textit{power-splitting} and \textit{time-sharing}. Among the proposed scenarios, the one that fits more in the requirements of the IoT devices is the joint design of downlink energy transfer and uplink information transfer, which is coordinated by the \textit{harvest-then-transmit-protocol}, according to which the devices first harvest energy, and then they transmit their independent messages to the access point.  Note that this setup is not dependent on the batteries capacity, since the harvested energy can be used directly or may charge super capacitors, which, in turn may replace or be combined with batteries, in order to ensure reliability. Recent scientific results have indicated that performance can be improved, while retaining the complexity at the devises' side,  when utilizing sophisticated schemes such as multiple antennas at the power beacon, or uplink NOMA \cite{NOMA_WPT_joutnal}. Thus, the utilization of SWIPT in IoT devices introduces non-trivial trade-offs between:
\begin{enumerate}
\item  the time dedicated for energy harvesting and that for information transmission; 
\item performance and complexity, 
\item performance and flexibility to diverse QoS requirements, and 
\item QoS and energy efficiency. 
\end{enumerate}
Finally, research challenges on this domain include standardization of evaluation models in the context of LPWANs.

\subsection{Security}
A basic principle of wireless communications is ``being connected is great, unless  you get exposed while poorly protected''. The same principle is valid for several IoT applications.  However, due to cost and energy considerations, unlicensed LPWANs does not offer any subscription identity module (SIM) authentication technique. Furthermore, LoRA and UNB encrypt neither the application payload nor the network joint request, i.e., they are exposed to eavesdropping~\cite{C:Low_trhoughput_networks_for_IoT}. 
Therefore, further study of authentication, security, and privacy techniques for LPWANs are~needed. 

\section{Conclusions}~\label{S:Conclusions}

IoT changes the requirements for connectivity significantly, mainly with regards to long battery life, low device costs, low deployment costs, extended coverage and support for a massive number of devices. Based on these requirements, several different cellular and non-cellular LPWAN solutions are emerging and are competing for IoT business and the overall connectivity market. 
In this paper, we reviewed the design specifications of different LPWANs solutions, as well as the demands of different IoT applications. 
Next,  we revisited the IoT applications and, based on the specifications of the LPWANs, we discuss the suitability of each LPWAN to meet the requirements of each IoT application. Finally, we presented the challenges  for LPWANs and we provided research directions.  
\bibliographystyle{IEEEtran}

\end{document}